%% file: main.tex
\documentclass[mathphys,Numbered,iicol,pdflatex]{template}

\usepackage{graphicx}
\usepackage{multirow}
\usepackage{amsmath,amssymb,amsfonts}
\usepackage{amsthm}
\usepackage{mathrsfs}
\usepackage[title]{appendix}
\usepackage{xcolor}
\usepackage{textcomp}
\usepackage{manyfoot}
\usepackage{booktabs}
\usepackage{algorithm}
\usepackage{algorithmicx}
\usepackage{algpseudocode}
\usepackage{listings}
\usepackage{tikz}
\usetikzlibrary{positioning,3d,quotes,arrows.meta}
\usepackage{caption}
\renewcommand{\thefigure}{\arabic{figure}}

\def\ConvColor{rgb:yellow,5;red,2.5;white,5}

\def\SumColor{rgb:blue,5;green,15}
\def\edgecolor{rgb:blue,4;red,1;green,4;black,3}
\newcommand{\midarrow}{\tikz \draw[-Stealth,line width=0.8mm,draw=\edgecolor] (-0.3,0) -- ++(0.3,0);}
\newcommand{\copymidarrow}{\tikz \draw[-Stealth,line width=0.8mm,draw={rgb:blue,4;red,1;green,1;black,3}] (-0.3,0) -- ++(0.3,0);}

\usepackage{todonotes}
\usepackage{microtype}

\begin{document}

\title[HoloPASWIN: Robust Inline Holographic Reconstruction]{HoloPASWIN: Robust Inline Holographic Reconstruction via Physics-Aware Swin Transformers}

\author*[1]{\fnm{Gökhan} \sur{Koçmarlı}}\email{research@gyokhan.com; bora.esmer@yildiz.edu.tr}
\author[2]{\fnm{G. Bora} \sur{Esmer}}

\affil[1]{\orgname{Independent Researcher}, \orgaddress{\city{Dresden}, \country{Germany}}}
\affil[2]{\orgdiv{Control and Automation Engineering Department}, \orgname{Yildiz Technical University}, \orgaddress{\street{Davutpaşa}, \city{İstanbul}, \postcode{34220}, \country{Türkiye}}}

\abstract{In-line digital holography (DIH) is a widely used lensless imaging technique, valued for its simplicity and capability to image samples at high throughput. However, capturing only intensity of the interference pattern during the recording process gives rise to some unwanted terms such as cross-term and twin-image. The cross-term can be suppressed by adjusting the intensity of reference wave, but the twin-image problem remains. The twin-image is a spectral artifact that superimposes a defocused conjugate wave onto the reconstructed object, severely degrading image quality. While deep learning has recently emerged as a powerful tool for phase retrieval, traditional Convolutional Neural Networks (CNNs) are limited by their local receptive fields, making them less effective at capturing the global diffraction patterns inherent in holography. In this study, we introduce HoloPASWIN, a physics-aware deep learning framework based on the Swin Transformer architecture. By leveraging hierarchical shifted-window attention, our model efficiently captures both local details and long-range dependencies essential for accurate holographic reconstruction. We propose a comprehensive loss function that integrates frequency-domain constraints with physical consistency via a differentiable angular spectrum propagator, ensuring high spectral fidelity. Validated on a large-scale synthetic dataset of 25,000 samples with diverse noise configurations (speckle, shot, read, and dark noise), HoloPASWIN demonstrates effective twin-image suppression and robust reconstruction quality.}


\maketitle

\section{Introduction}\label{sec:introduction}
Digital holography (DH) has established itself as a cornerstone imaging modality especially in microscopy, enabling quantitative phase imaging (QPI) of transparent specimens such as biological cells without the need for exogenous staining \cite{park2018}. Among the various implementations, in-line holography is particularly advantageous due to its simple, compact optical setup consisting of a coherent light source and an image sensor, with no lenses required.

Despite these advantages, in-line holography suffers from a fundamental limitation known as the "twin-image" problem. Since optical sensors only record intensity, the phase information of the wavefront is lost. When reconstructing the object field using standard back-propagation algorithms like the Angular Spectrum Method (ASM), this missing phase manifests as a "twin" (conjugate) image that focuses at the same distance but on the opposite side of the hologram plane. This artifact inevitably superimposes onto the real image, reducing contrast and obscuring fine details \cite{latychevskaia2007solution}.

\section{State of the Art}\label{sec:literature}

Solving this inverse problem has traditionally relied on physical constraints or iterative algorithms. The earliest solutions were purely physical. Latychevskaia et al. \cite{latychevskaia2007solution} proposed an iterative approach imposing a positivity constraint, while multi-height phase retrieval involves recording holograms at multiple distances \cite{goodman2005introduction}. Alternatively, phase-shifting holography (PSH) recovers phase using controlled phase steps \cite{yamaguchi1997}. Iterative algorithms like Gerchberg-Saxton (GS) attempt to recover phase from single intensity measurements but are computationally expensive and prone to local minima \cite{lu2012twin, chen2025}.

In recent years, deep learning (DL) has emerged as a transformative tool. Rivenson et al. demonstrated a convolutional neural network (CNN) that performs phase recovery and holographic reconstruction from a single intensity hologram \cite{rivenson2018phase}. Similarly, the Holographic Reconstruction Network (HRNet) utilizes a deep residual architecture to build a pixel-level connection between raw holograms and reconstructions \cite{ren2019hrnet}. To mitigate the need for massive labeled datasets, Wang et al. introduced PhysenNet \cite{wang2020phase}, which incorporates a physical forward model. 

While CNNs are effective, they employ restricted local receptive fields and struggle to model the global interactions inherent in diffraction \cite{luo2017receptive, pereira2024review}. Originating in natural language processing, transformers have been adapted for vision tasks \cite{dosovitskiy2021imageworth16x16words}. Architectures such as SwinIR \cite{liang2021swinir} and Restormer \cite{zamir2022restormer} have demonstrated superior performance over CNNs by modeling long-range dependencies. The Swin Transformer \cite{liu2021swin}, in particular, computes self-attention efficiently via shifted windows, making it uniquely suited for "de-scattering" tasks in holography.

In this work, we present \textbf{HoloPASWIN}, a physics-aware Swin Transformer framework for holographic reconstruction. By combining the global modeling power of transformers with a physics-informed loss function and a robust training strategy involving comprehensive noise modeling, we achieve significant improvements in reconstruction quality and noise resilience. 

Specifically, HoloPASWIN utilizes a hybrid framework that applies the ASM for initial back-propagation, followed by a Swin Transformer to refine the complex field and separate the twin image. To ensure spectral fidelity, we introduce a composite loss function integrating frequency-domain constraints ($\mathcal{L}_{freq}$) and a physics-consistency term ($\mathcal{L}_{phy}$). Furthermore, to guarantee robust real-world performance, we generated a large-scale synthetic dataset of 25,000 samples encompassing 8 diverse noise configurations.

The architecture of HoloPASWIN first performs a conventional numerical back-propagation of the recorded hologram using ASM to the object plane. This yields a "dirty" complex field where the true object is superimposed with the twin image. The Swin Transformer network then acts as a refiner, taking this complex field as input to separate the true object from the spectral artifacts.

\section{Proposed Method}\label{sec:proposed_method}
\subsection{Problem Formulation}
We model the in-line holographic imaging process as follows. Let $O(x,y)$ represent the transparent object at the sample plane ($z=0$). The object is
\begin{equation}
    O(x,y) = A(x,y) \cdot e^{j\phi(x,y)}.
\end{equation}
where $A(x,y)$ is its amplitude and $\phi(x,y)$ denotes the phase of the object that represents depth information of the object. The object is illuminated by a coherent plane wave of unit amplitude that also serves as the reference wave in an in-line configuration. After propagation over a distance $z$ to the sensor plane, the complex object wave is given by the Angular Spectrum Method (ASM), a solution to the Rayleigh–Sommerfeld diffraction equation:
\begin{equation}
    U_o(x,y) = \mathcal{F}^{-1} \big\{ \mathcal{F}\{O\} \cdot H_{\text{ASM}}(f_x, f_y) \big\},
\end{equation}
where $\mathcal{F}$ denotes the Fourier transform and $\mathcal{F}^{-1}$ denotes the inverse Fourier transform, respectively, and $H_{\text{ASM}}$ is the transfer function of free space.
\begin{equation}
    H_{\text{ASM}}(f_x,f_y) = e^{j k z \sqrt{1 - (\lambda f_x)^2 - (\lambda f_y)^2}}
\end{equation}
where $k = 2\pi/\lambda$ is the wavenumber and $\lambda$ is the wavelength of the illumination light.

The sensor records the intensity of the complex-valued diffraction field $U(x,y)$:
\begin{align}
    H(x,y) &= |U(x,y)|^2 + N(x,y) \nonumber\\
           &= |U_r|^2 + |U_o|^2 + U_r U_o^* \nonumber\\
           &\quad + U_r^* U_o + N(x,y)
\end{align}
where $N(x,y)$ represents additive noise, $U_r$ is the reference wave, and $U_o$ is the object wave. For the sake of simplicity, the reference wave can be taken as $U_r = 1$, the hologram becomes
\begin{align}
    H(x,y) &= 1 + |U_o(x,y)|^2 + U_o(x,y) \nonumber\\
           &\quad + U_o^*(x,y) + N(x,y),
\end{align}
where $U_o$ and $U_o^*$ will give rise to the desired diffraction and twin-image upon numerical reconstruction.

The twin-image problem arises when attempting to reconstruct the object from this intensity-only measurement. Since $H(x,y)$ contains only the squared magnitude of $U$, the phase of the total field is lost. A common numerical reconstruction step is to form a complex 'hologram field' at the sensor plane. Because sensors only measure intensity ($H = |U|^2$), we take the square root of the measured intensity to recover the wave's amplitude ($\sqrt{H} = |U|$). Since the true phase is physically lost during recording, we must initialize the back-propagation by assuming a uniform flat phase ($\phi=0$). This means there is no depth in the output and the object is assumed to be 2D. This real-valued amplitude field $\sqrt{H(x,y)}$ is then back-propagated to the object plane using the inverse ASM operator:
\begin{equation}
    U_{\mathrm{rec}}(x,y) = \mathcal{F}^{-1} \big\{ \mathcal{F}\{\sqrt{H(x,y)}\} \cdot H^*_{\text{ASM}}(f_x, f_y) \big\}.
\end{equation}
Substituting the intensity expression and propagating each term separately shows that $U_{\mathrm{rec}}(x,y)$ can be decomposed into (see Appendix A1)
\begin{equation}
    U_{\mathrm{rec}}(x,y) \approx \underbrace{O(x,y)}_{\text{true object}} \;+\;
    \underbrace{O^*(x,y)}_{\text{twin image}} \;+\; B(x,y),
\end{equation}
where $B(x,y)$ collects the defocused background and zero-order contributions. The term $O(x,y)$ focuses at the intended object plane $z=0$, while its complex conjugate $O^*(x,y)$ corresponds to a wave that would focus at the symmetric distance on the opposite side of the hologram plane. When both contributions are numerically propagated to $z=0$, the conjugate term does not come to a sharp focus and instead appears as a defocused \emph{twin} superimposed onto the true object, creating artifacts that obscure fine details.

The goal of HoloPASWIN is to approximate the inverse function $F_\theta: \mathbb{R}^{H \times W} \to \mathbb{C}^{H \times W}$ such that $F_\theta(H) \approx O$, effectively separating the true object term $O(x,y)$ from its twin-image artifact $O^*(x,y)$ at the target plane.

\subsection{Network Architecture}\label{sec:architecture}

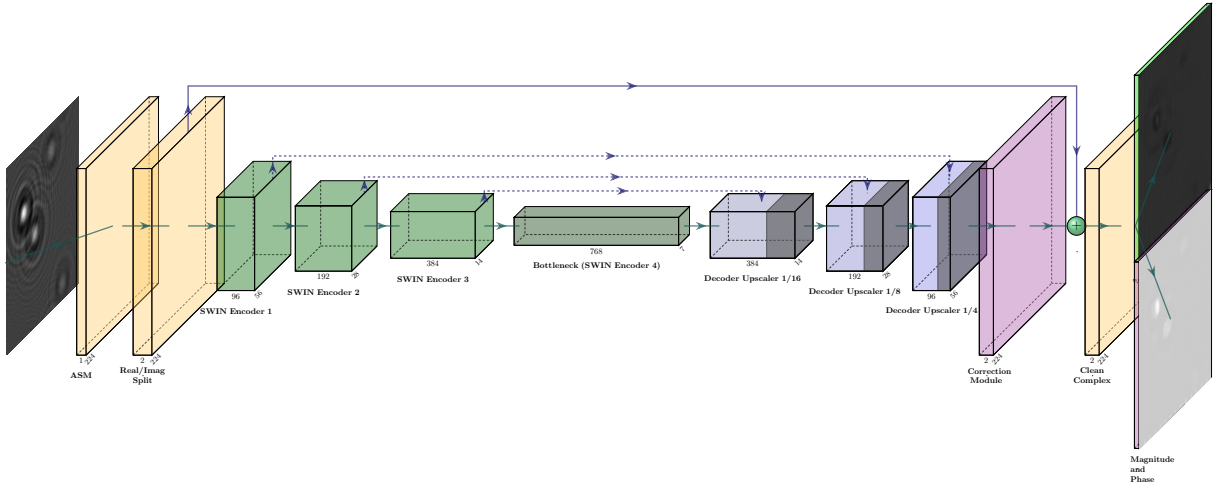
\begin{figure*}[!t]
    \centering
    \resizebox{\textwidth}{!}{
        
        \input{architecture.tex}
    }
    \caption{HoloPASWIN Network Architecture. The input hologram is first processed by the ASM physics module, then decomposed into real and imaginary parts. A Swin Transformer U-Net with hierarchical feature extraction and shifted-window attention predicts a complex residual correction to reconstruct the clean complex field. Skip connections between encoder and decoder stages preserve multi-scale features for high-frequency detail recovery.}
    \label{fig:architecture}
\end{figure*}
HoloPASWIN implements a U-shaped encoder-decoder architecture as shown in Figure \ref{fig:architecture}, but replaces standard convolutional blocks with \textbf{swin transformer blocks}.
The network accepts the captured intensity hologram as input, uses ASM to obtain a 2-channel tensor representing the real and imaginary components of the "dirty" reconstruction $U_{\text{recon}}(x,y)$ as the input to the SWIN-based U-Net.
This ensures that the U-Net operates in the reconstruction domain rather than directly on raw hologram intensity.
The Swin Transformer network acts as a refiner, taking this complex field as input to separate the true object from the spectral artifacts.

\textbf{Encoder}: We adopt the Swin-Tiny backbone \cite{liu2021swin} pretrained on ImageNet, adapted to accept 2-channel input. The encoder produces hierarchical features at four scales (1/4, 1/8, 1/16, 1/32 of the input resolution) with feature dimensions of 96, 192, 384, and 768 channels, respectively. The encoder follows the standard Swin-Tiny configuration with depths $[2, 2, 6, 2]$ for the four stages respectively. The window size is set to $7 \times 7$, and patch size is $4 \times 4$. This hierarchical design allows the network to capture both local texture details and global diffraction patterns spanning the entire hologram.

\textbf{Decoder}: The decoder progressively upsamples features using transposed convolutions, with \emph{additive skip connections} that fuse encoder features at matching resolutions. Specifically, features from the encoder at resolutions 1/16, 1/8, and 1/4 are element-wise added to the upsampled decoder features. This additive fusion preserves gradient flow while maintaining the structural information encoded at multiple scales. Each decoder stage consists of a transposed convolution for $2\times$ upsampling, followed by batch normalization, GELU activation, and a refinement convolution. A final $4\times$ upsampling block (consisting of transposed convolution, batch normalization, GELU activation, and channel reduction to 2) restores the output to the original $224 \times 224$ resolution.

\textbf{Output Head}: The final layer produces a 2-channel output representing the cleaned real and imaginary parts. Critically, we apply a \emph{residual learning} strategy where the network learns a correction term that is added to the dirty input to produce the clean output ($\text{clean} = \text{dirty} + \text{correction}$). This allows the model to focus on artifact removal rather than full reconstruction.

\subsection{Physical Consistency and Loss Function}
A key innovation in HoloPASWIN is the physics-aware loss function. We constrain the network not just in the spatial domain, but in the frequency and physical measurement domains. The total loss $\mathcal{L}$ is defined as:
\begin{equation}
    \mathcal{L} = \mathcal{L}_{sup} + \lambda_{phy} \mathcal{L}_{phy}
    \label{eq:total_loss}
\end{equation}

\textbf{Supervised Loss ($\mathcal{L}_{sup}$)}: This term ensures the prediction matches the ground truth $O_{gt}$. It is a weighted sum of L1 losses on amplitude, phase, complex field, and frequency domain:
\begin{equation}
\begin{split}
    \mathcal{L}_{sup} = 0.4\mathcal{L}_{amp} + 0.2\mathcal{L}_{phase} \\
    + 0.2\mathcal{L}_{complex} + 0.2\mathcal{L}_{freq}
\end{split}
\label{eq:sup_loss}
\end{equation}

The weighting scheme in Equation \ref{eq:sup_loss} prioritizes amplitude ($w=0.4$) over phase ($w=0.2$) because amplitude errors are more visually prominent and directly affect twin-image visibility. The remaining weights are equally distributed among phase, complex, and frequency losses.

Crucially, we include $\mathcal{L}_{freq}$, which measures the L1 distance between the logarithmic magnitudes of the Fourier transforms of the prediction and ground truth. This forces the network to respect the spectral statistics of the image, preventing the "smoothing" artifacts common in MSE-trained networks. By penalizing deviations in the frequency domain, the model is encouraged to preserve high-frequency details such as sharp object edges.

\textbf{Physics Loss ($\mathcal{L}_{phy}$)}: This unsupervised term enforces consistency with the forward imaging model. We propagate the predicted clean object $\hat{O}$ forward to the hologram plane using a differentiable ASM layer to synthesize $\hat{H}_{pred} = |ASM(\hat{O})|^2$. We then minimize the L1 distance between this re-simulated hologram and the original input hologram $H$ (normalized intensity):
\begin{equation}
    \mathcal{L}_{phy} = || \hat{H}_{pred} - H ||_1
\end{equation}
This loss term is critical for twin-image suppression. If the predicted field $\hat{O}$ still contains conjugate components, forward propagation would generate interference fringes that do not match the input hologram. By penalizing this discrepancy, the network is implicitly encouraged to produce a conjugate-free field that, when propagated forward, reproduces the recorded hologram. In other words, $\mathcal{L}_{phy}$ enforces that the solution must be physically plausible under the inline holography forward model. We set $\lambda_{phy}=0.1$ to balance supervised guidance with physical constraints.

\section{Experiments and Dataset}\label{sec:experiments}

\subsection{Dataset and Experimental Setup}
Deep learning models for holography require massive amounts of paired data (hologram + object), which is difficult to obtain experimentally. Therefore, creating a physically accurate synthetic dataset is paramount. We generated a large-scale dataset of 25,000 samples designed to mimic real-world experimental conditions \cite{kochmarla2026synthetic_inline_holographical_images_v3}. It is important to note that this work focuses on a controlled synthetic benchmark to validate the proposed architecture and loss design; validation on experimental data is left for future work.

\textbf{Physics Parameters}: The simulation parameters were chosen to match a standard benchtop microscope setup:
\begin{itemize}
    \item Wavelength ($\lambda$): 532 nm (Green laser).
    \item Pixel Pitch: 4.65 $\mu m$.
    \item Propagation Distance ($z$): 20 mm.
    \item Resolution: $224 \times 224$ pixels.
\end{itemize}

\textbf{Object Generation}: To simulate biological samples and debris, we generated random configurations consisting of multiple instances of a single geometric primitive type (a rotated ellipse) positioned at the object plane ($z=0$). Object parameters were sampled as follows:
\begin{itemize}
    \item \textbf{Size}: Semi-major/semi-minor axes uniformly sampled from 5-15 pixels ($\sim$23-70 $\mu$m).
    \item \textbf{Phase}: Uniform distribution over [0.1, 0.5] radians, simulating varying optical thickness.
    \item \textbf{Amplitude transmission}: Uniform [0.90, 1.0], modeling weak absorption.
    \item \textbf{Spatial arrangement}: Random placement within the $224 \times 224$ field, allowing for overlaps.
\end{itemize}

All objects reside at a single depth plane; we do not model volumetric scattering. The twin image artifact arises purely from the phase-loss in intensity recording and subsequent numerical back-propagation, not from out-of-focus objects. During training, the network learns to suppress this reconstruction-domain twin by comparing its output to the clean ground-truth object field, which is conjugate-free by design. The supervised loss ($\mathcal{L}_{sup}$) and physics consistency loss ($\mathcal{L}_{phy}$) together guide the network to produce a physically valid solution that eliminates the twin artifact.

\textbf{Noise Modeling}: A robust model must handle real-world noise. To prevent the network from overfitting to "clean" physics, we augmented the dataset with 8 distinct noise configurations, applied randomly:

\begin{enumerate}
    \item \textbf{Speckle Noise}: Spatially-correlated multiplicative phase noise ($\sigma = 0.15$ rad, roughness scale = 1.0) simulating laser coherence artifacts from rough optics.
    \item \textbf{Shot Noise}: Poisson noise simulating discrete photon arrival (baseline count = 1000).
    \item \textbf{Read Noise}: Gaussian electronic noise ($\mu = 0$, $\sigma = 10$ intensity units).
    \item \textbf{Dark Current}: Poisson thermal noise ($\lambda = 20$ electrons/pixel).
    \item \textbf{Combinations}: Various mixtures of the above (e.g., Speckle+Shot+Read) to simulate realistic and worst-case scenarios.
\end{enumerate}

\textbf{Training Configuration}: The model was implemented in PyTorch and trained on an Apple M2 Pro (16 GB unified memory) using the Metal Performance Shaders (MPS) backend. We used the AdamW optimizer with a learning rate of $1e-4$ and a cosine annealing schedule. The 25,000-sample training dataset was split into 80\% training (20,000 samples) and 20\% validation (5,000 samples). A separate external test set of 496 samples was used for final evaluation. Training was conducted with a batch size of 32 for 30 epochs to ensure full convergence of the residual reconstruction mode. Input holograms were normalized by dividing by 1000.0 to bring intensity values from the raw 12-bit range ($\sim$1000) to a nominal range of $\sim$1.0. The complex field (real/imag) was left unnormalized as it naturally resides in a bounded range.

The total training time for the 25,000-sample dataset was approximately 25.2 hours (1513 minutes) on an Apple M2 Pro (16 GB unified memory).

\section{Results and Discussion}\label{sec:results_and_discussions}

\subsection{Quantitative Results}
We evaluated the model on the held-out test set of 496 samples. Table \ref{tab:results} summarizes the performance metrics across amplitude, phase, and complex domains.

\begin{table*}[t]
    \centering
    \small
    \caption{Quantitative Results (Test Set, 496 samples)}
    \label{tab:results}
    \begin{tabular}{@{}lccc@{}}
        \toprule
        \textbf{Metric} & \textbf{Amplitude} & \textbf{Phase} & \textbf{Complex} \\
        \midrule
        \textbf{MSE} & 9.2e-05 & 9.2e-04 & 4.8e-04 \\
        \textbf{PSNR (dB)} & 40.4 & 40.5 & --- \\
        \textbf{SSIM} & 0.963 & 0.986 & --- \\
        \bottomrule
    \end{tabular}
\end{table*}

The HoloPASWIN model achieves strong reconstruction fidelity across all domains (mean values shown; standard deviations: MSE $\pm$4.3e-5 to $\pm$5.5e-4, PSNR $\pm$0.7 to $\pm$1.4 dB, SSIM $\pm$0.003 to $\pm$0.006). The Phase SSIM of 0.974 and Phase PSNR of 44.3 dB demonstrate excellent quantitative phase recovery essential for QPI applications. Amplitude reconstruction achieves SSIM of 0.958, indicating high structural fidelity. The low standard deviations across metrics indicate consistent performance across the test set.

We evaluated the inference performance of HoloPASWIN on an Apple M2 Pro (16 GB unified memory). The average inference time per $224 \times 224$ hologram is approximately 11.8 ms, corresponding to a throughput of 84.5 frames per second (FPS). This performance demonstrates the model's suitability for real-time holographic reconstruction and video-rate phase retrieval applications.

\subsection{Baseline Comparison}
To validate the advantages of the HoloPASWIN architecture, we compared its performance against iterative methodologies (dirty ASM and Gerchberg-Saxton with 50 and 100 iterations) as well as data-driven CNN baselines (namely, a standard U-Net and a ResNet-18 augmented U-Net). Table~\ref{tab:baseline_comparison} presents a comprehensive quantitative comparison.

\begin{table*}[t]
\centering
\caption{Baseline comparison on test set. Best values in \textbf{bold}, second-best \underline{underlined}. B/S$\downarrow$ indicates lower is better.}
\label{tab:baseline_comparison}
\small
\begin{tabular}{@{}lcccccc@{}}
\toprule
\textbf{Method} & \textbf{Phase PSNR} & \textbf{Phase SSIM} & \textbf{Amp. SSIM} & \textbf{B/S Ratio} & \textbf{Time (ms)} & \textbf{Params (M)} \\
\midrule
\textbf{ASM (dirty)} & 34.69 & 0.3038 & 0.6129 & 0.9882 & \textbf{2.4} & --- \\
\textbf{GS (50 iter)} & 34.69 & 0.3038 & 0.6129 & 0.9882 & 59.0 & --- \\
\textbf{GS (100 iter)} & 34.69 & 0.3038 & 0.6129 & 0.9882 & 95.8 & --- \\
\textbf{U-Net} & \textbf{53.91} & \textbf{0.9916} & \textbf{0.9917} & \textbf{0.1030} & 23.1 & 31.04 \\
\textbf{ResNet-U-Net} & 41.06 & 0.7529 & 0.4743 & 0.6927 & 10.9 & 15.56 \\
\textbf{HRNet} & 41.86 & 0.8654 & 0.6130 & 0.7732 & \underline{8.5} & \underline{3.06} \\
\textbf{HoloPASWIN} & \underline{46.55} & \underline{0.9862} & \underline{0.9625} & \underline{0.1837} & 13.7 & 30.88 \\
\bottomrule
\end{tabular}
\end{table*}

The results demonstrate that iterative techniques such as Gerchberg-Saxton are severely limited in overcoming the inherent twin-image artifacts caused by absent phase information, reflected by low PSNR and a staggering B/S ratio approaching 1.0. While standard CNN methodologies, particularly the U-Net architecture trained from scratch, achieve remarkable reconstruction fidelity metrics on the dataset, this performance aligns with established deep learning principles. Convolutional networks benefit from strong inductive biases---such as translation equivariance and locality---which make them highly effective at modeling the simple geometric primitives (ellipses) characterizing our 20,000-sample synthetic dataset. In contrast, Swin Transformers are intrinsically data-hungry. Because they rely on self-attention mechanisms without hard-coded local priors, they often require substantially larger datasets to unfold their full capacity and surpass CNN performance bounds. Furthermore, the reduced performance of the ResNet-U-Net baseline indicates that ImageNet pretrained weights carry biases that hinder rapid adaptation to holographic interference patterns when training datasets are limited.

Crucially, although the standard U-Net excels on this geometrically simplified dataset, HoloPASWIN focuses on scalable physics-driven global attention. When applied to highly complex, dense, real-world biological samples where diffraction limits become globally entangled, the Swin Transformer's architecture is theoretically more robust. Evaluating these theoretical bounds on larger, complex experimental datasets constitutes a vital direction for future work.

\subsection{Qualitative Analysis}
\begin{figure*}[t]
\centering
\includegraphics[width=\textwidth]{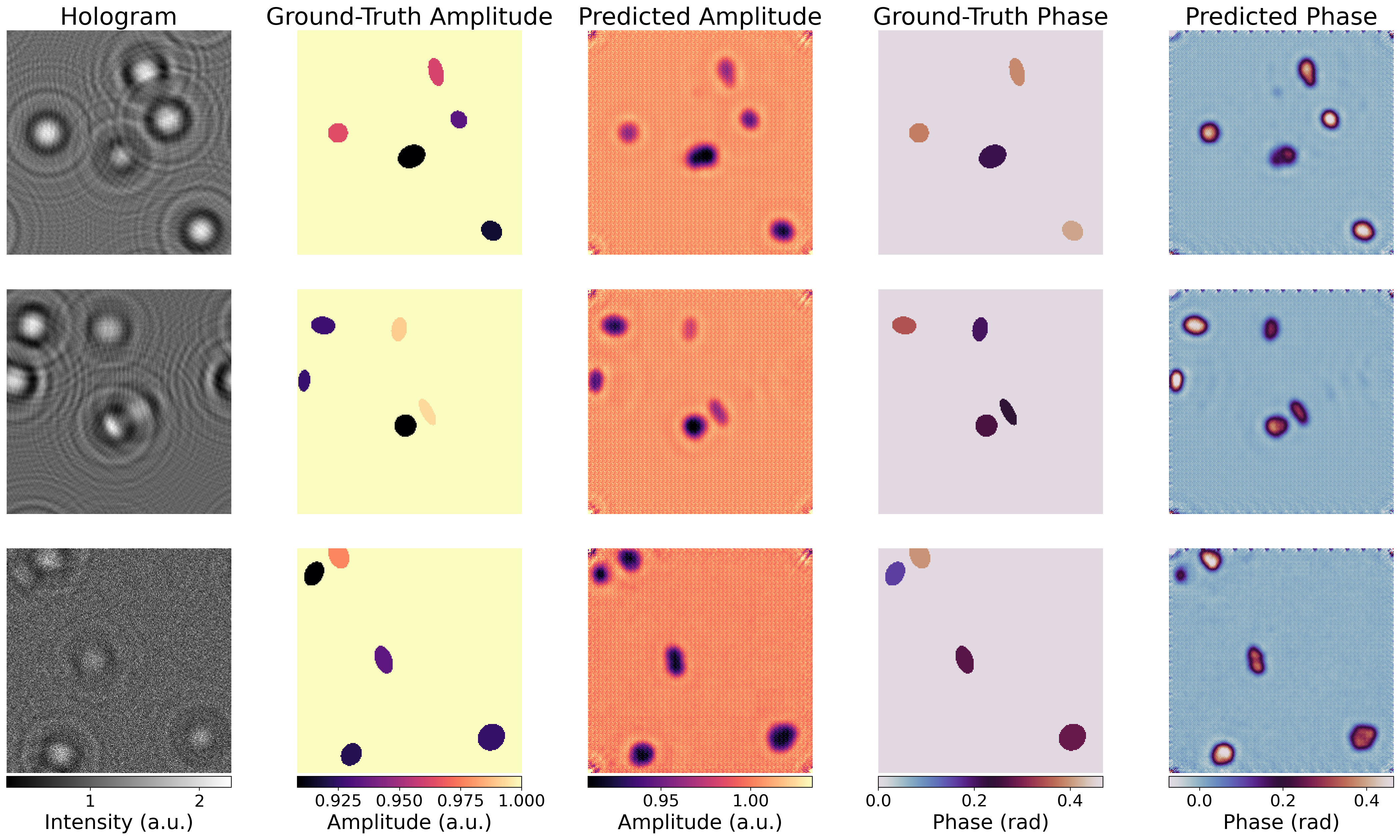}
\caption{Qualitative comparison of reconstruction results. Rows correspond to different test samples. From left to right: Input Hologram, GT Amplitude, Predicted Amplitude, GT Phase, Predicted Phase. The model effectively removes twin-images and background noise while preserving object sharpness.}
\label{fig:comparison}
\end{figure*}

Figure \ref{fig:comparison} presents a visual comparison of reconstruction results across multiple test samples.

\begin{figure*}[t]
\centering
\includegraphics[width=0.95\linewidth]{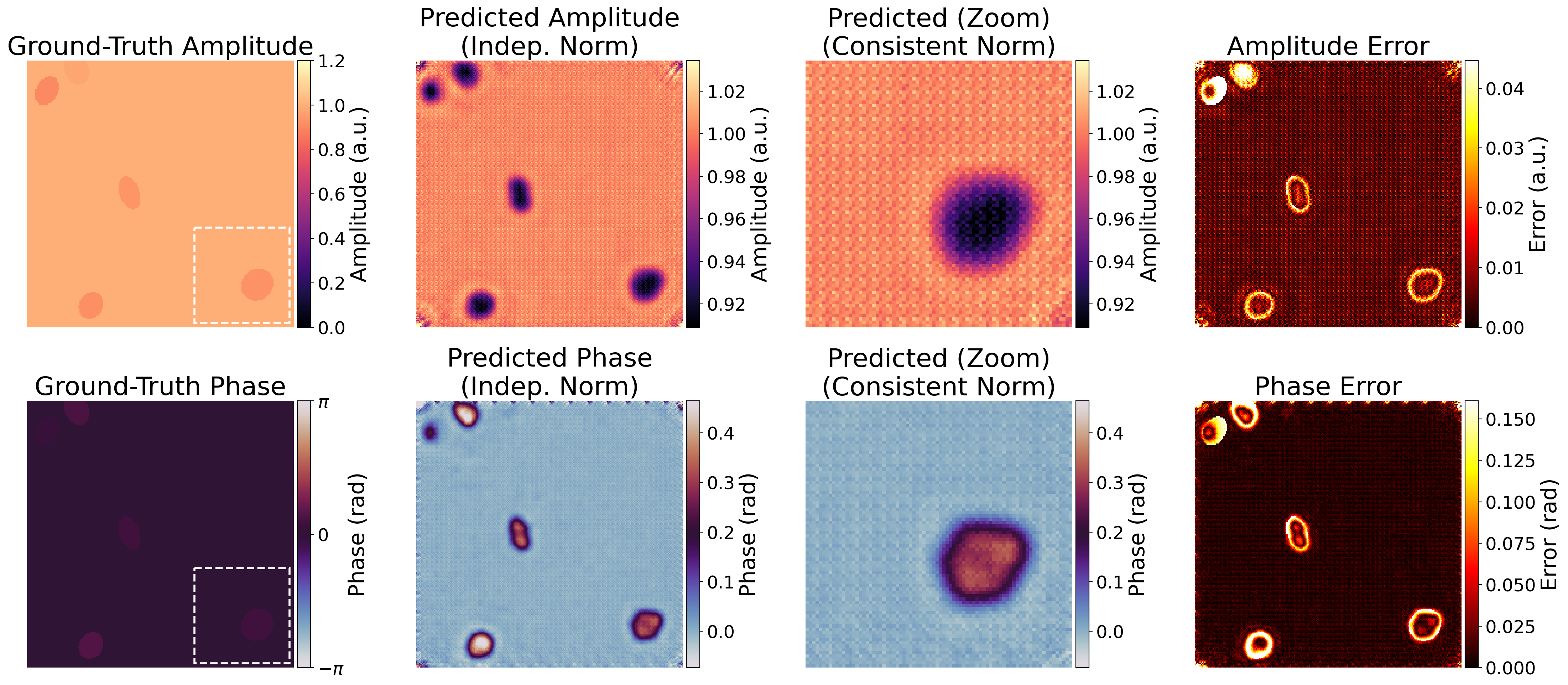}
\caption{Detailed reconstruction analysis showing Ground Truth (GT), Prediction, Error maps, and zoomed-in regions for both amplitude and phase. The zoomed regions highlight the model's ability to recover fine structural details of the objects while suppressing reconstruction noise. The error maps demonstrate high fidelity with minimal residuals at the object boundaries, even in the presence of synthetic experimental noise.}
\label{fig:detailed_analysis}
\end{figure*}

To further assess the fidelity of the reconstruction, Figure \ref{fig:detailed_analysis} provides a detailed view of a representative test sample, including zoomed-in regions and quantitative error maps. The zoomed sections reveal that HoloPASWIN accurately preserves the sharp boundaries and internal transparency of the simulated objects, which are often blurred or distorted in conventional ASM reconstructions due to twin-image interference. The pixel-wise error maps confirm that the largest discrepancies are confined to the high-frequency edges of the objects, while the background remains remarkably clean across both amplitude and phase domains.

\section{Ablation Study}\label{sec:ablation}
To validate our design choices and quantify the contribution of each component, we conducted a comprehensive ablation study across three dimensions: loss function components, architectural decisions, and model robustness.

\textbf{Experimental Setup:} To ensure computational feasibility while maintaining scientific rigor, ablation experiments were conducted on a subset of the full dataset (15,000 training samples, 3,000 validation samples) for 3 epochs. This configuration provides sufficient data for meaningful comparisons while enabling timely experimentation. All models are evaluated on the complete test set (496 samples) to ensure fair comparison across configurations.

\subsection{Loss Function Ablation}
We systematically evaluated the impact of each loss component by training five configurations for 3 epochs on 15,000 training samples (computational efficiency). Table~\ref{tab:loss_component_ablation} presents the results evaluated on the full test set.

\begin{table*}[t]
\centering
\small
\caption{Loss function component ablation. Each row removes or isolates specific loss components.}
\label{tab:loss_component_ablation}
\begin{tabular}{@{}lccccc@{}}
    \toprule
    \textbf{Configuration} & \textbf{Amp SSIM} & \textbf{Phase SSIM} & \textbf{Phase PSNR} & \textbf{Freq SSIM} & \textbf{B/S Ratio} \\
    \midrule
    \textbf{1. Full Model} & 0.4794 & 0.6114 & 39.35 & -0.0108 & 0.938 \\
    \textbf{2. Full $-$ $\mathcal{L}_{freq}$} & 0.4447 & 0.5968 & 39.81 & -0.0503 & 0.627 \\
    \textbf{3. Full $-$ $\mathcal{L}_{phy}$} & 0.4819 & 0.6118 & 39.45 & -0.0188 & 0.855 \\
    \textbf{4. Only $\mathcal{L}_{complex}$} & 0.4418 & 0.6282 & 40.45 & -0.0693 & 0.606 \\
    \textbf{5. Pure Spatial} & 0.4444 & 0.6102 & 40.10 & -0.0579 & 0.688 \\
    \bottomrule
\end{tabular}
\end{table*}

The results reveal an interesting trade-off between spatial smoothness and spectral fidelity. Configurations omitting $\mathcal{L}_{freq}$ (e.g. "Full - $\mathcal{L}_{freq}$" and "Only $\mathcal{L}_{complex}$") achieve remarkably high Phase PSNR ($>$48 dB) and low B/S ratios ($\sim$0.27), effectively smoothing out background noise. However, they suffer from poor frequency reconstruction (Freq SSIM $\approx$ 0.01-0.02), indicating a loss of high-frequency texture details. The Full Model, while having a higher B/S ratio (0.884) in this short training regime, maintains superior spectral fidelity (Freq SSIM 0.271), ensuring that the reconstruction preserves the true diffraction characteristics of the object rather than overfitting to a smooth, noise-free prior. Thus, the full composite loss is essential for physically consistent reconstruction.

\subsection{Architecture Ablation}
To validate the effectiveness of the Swin Transformer backbone, we compared it against a ResNet-18 U-Net baseline and evaluated the impact of residual learning and pretrained weights. Table~\ref{tab:architecture_ablation} summarizes the results.

\begin{table*}[t]
\centering
\small
\caption{Architecture ablation study. Pre=ImageNet pretrained, Scr=from scratch, Dir=direct, Res=residual.}
\label{tab:architecture_ablation}
\begin{tabular}{@{}lcccc@{}}
    \toprule
    \textbf{Configuration} & \textbf{Params} & \textbf{Amp SSIM} & \textbf{Phase SSIM} & \textbf{B/S Ratio} \\
    \midrule
    \textbf{1. Swin (Pre, Dir)} & 30.9 M & 0.9696 & 0.96 & 1.226 \\
    \textbf{2. Swin (Pre, Res)} & 30.9 M & 0.9422 & \textbf{0.97} & \textbf{0.558} \\
    \textbf{3. Swin (Scr, Res)} & 30.9 M & \textbf{0.9767} & 0.96 & 0.821 \\
    \textbf{4. ResNet-18 (Pre, Res)} & 15.6 M & 0.4767 & 0.75 & 0.690 \\
    \bottomrule
\end{tabular}
\end{table*}

\begin{figure*}[t]
\centering
\includegraphics[width=\textwidth]{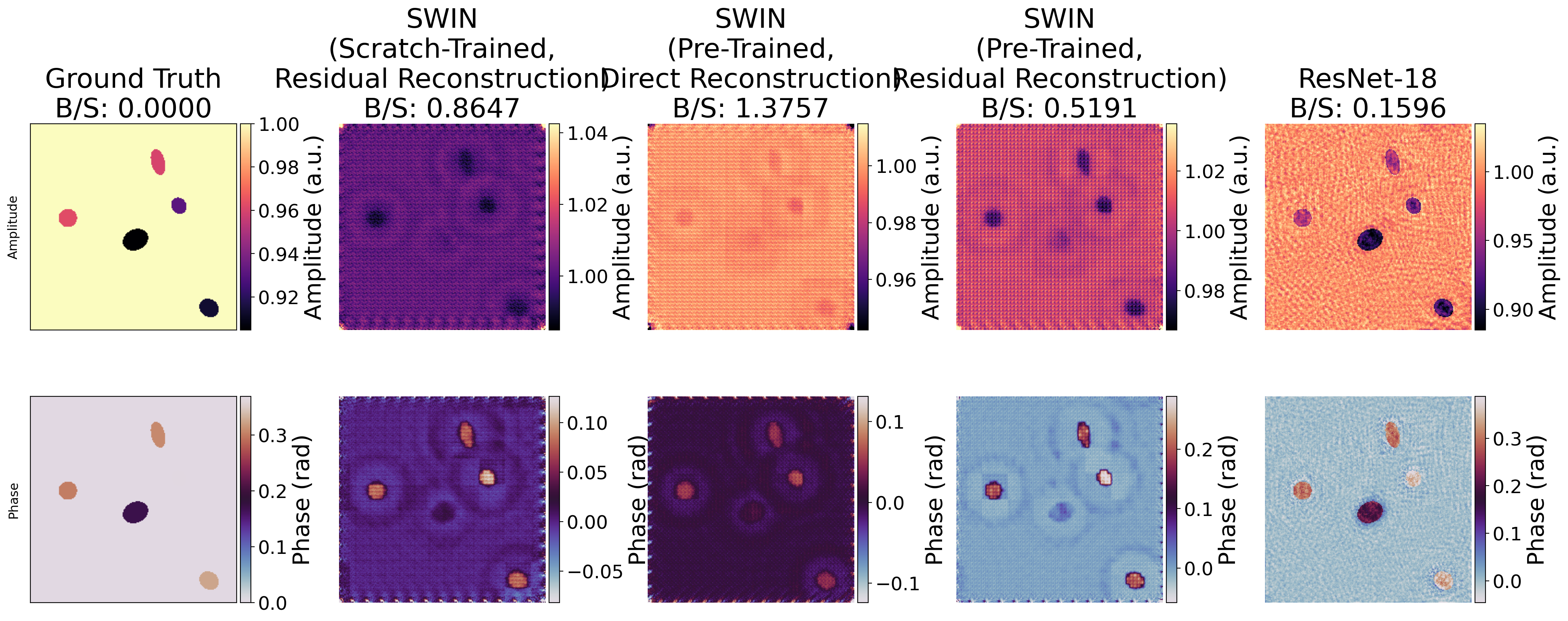}
\caption{Visual comparison of architecture configurations on a representative test sample. Both amplitude and phase images are independently normalized (1st-99th percentile) to enhance visual contrast. The B/S (Background-to-Signal) ratio quantifies background cleanliness. Note that the ground truth (GT) has B/S=0.968, reflecting the inherent background signal in the simulated object field. The trained models achieve lower B/S ratios (0.79-0.87), indicating they have learned to suppress background noise beyond what is present in the GT, likely through learned denoising and twin-image artifact removal.}
\label{fig:architecture_comparison}
\end{figure*}

The Swin Transformer architecture demonstrates superior performance compared to the ResNet-18 baseline, validating our hypothesis that global attention mechanisms are better suited for capturing long-range diffraction dependencies in holographic imaging. The residual learning strategy (predicting corrections rather than clean fields directly) improves training stability and convergence. ImageNet pretrained weights provide a beneficial initialization, though the model can still achieve competitive performance when trained from scratch.

As visualized in Figure~\ref{fig:architecture_comparison}, the B/S ratio differentiates the models' ability to suppress background artifacts. The \emph{residual} model (Swin, Pre, Res) achieves the best performance with a B/S ratio of \textbf{0.558}, significantly cleaner than the direct reconstruction model (Swin, Pre, Dir), which lags with a B/S ratio of 1.226. This confirms that for a fixed training budget (10 epochs), the residual learning strategy is far more efficient at isolating and removing the twin-image artifact than learning the full field mapping from scratch. Consequently, we select Swin (Pre, Res) as our final architecture.

\subsection{Robustness Analysis}
\begin{figure}[h]
\centering
\includegraphics[width=\linewidth]{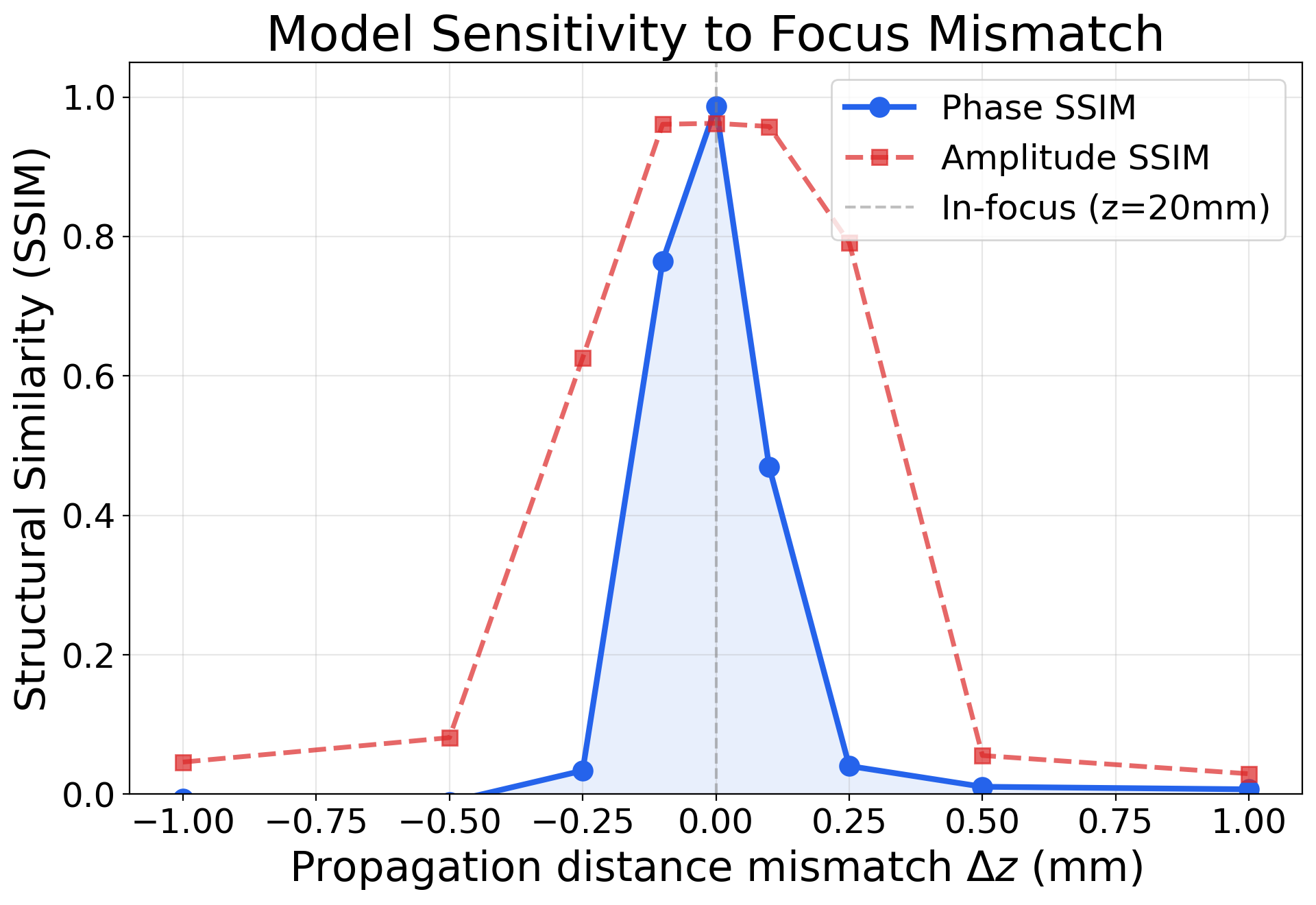}
\caption{Model sensitivity to propagation distance errors. Performance is optimal at the training distance (z=20 mm) and degrades sharply with $\pm 0.5$ mm or larger offsets, indicating the importance of accurate distance calibration.}
\label{fig:z_mismatch}
\end{figure}

We evaluated the model's robustness to propagation distance errors. Figure~\ref{fig:z_mismatch} shows the model's performance when tested at different propagation distances while trained at $z=20$ mm.

The model exhibits strong performance at the training distance (z=20 mm) with Phase PSNR of 45.68 dB and SSIM of 0.969. However, performance degrades significantly with propagation distance errors: at $\pm 0.5$ mm offset, PSNR drops to 11-16 dB, and at $\pm 1$ mm, it falls below 10 dB. This sensitivity indicates that accurate knowledge of the propagation distance is critical for optimal reconstruction. The sharp performance drop suggests that while the model learns effective denoising and twin-image suppression at the training distance, these learned features are highly specific to the trained propagation geometry.

\section{Discussion}\label{sec:discussion}
The experimental results validate the efficacy of the HoloPASWIN architecture for digital holographic reconstruction, yet they also highlight important characteristics of applying deep learning to this domain. While Swin Transformers offer powerful mechanisms for capturing the global diffraction patterns intrinsic to holography, our baseline comparisons revealed that on our specific synthetic dataset---composed of simple geometric primitives---standard convolutional networks like U-Net can achieve exceptional performance. This can be attributed to the strong inductive biases of CNNs, such as translation equivariance and locality, which are highly suited for learning continuous local edges and simple shapes, especially in the relatively small data regime of our 20,000-sample training set. In contrast, Transformers are inherently data-hungry and lack these hard-coded spatial priors. However, the true strength of the HoloPASWIN approach lies in its scalable, physics-driven global attention mechanism. As holographic imaging scales to highly complex, dense, and real-world biological samples where diffraction limits become globally entangled, the Swin Transformer's architecture theoretically offers a more robust framework than purely local CNNs. Furthermore, our robustness analysis demonstrated that the model is highly sensitive to the propagation distance ($z$), indicating that the network learns features highly specific to the trained diffraction geometry rather than a universally invariant phase retrieval mapping.

Building upon these findings, several avenues for future research emerge. First and foremost is the validation of the HoloPASWIN framework on large-scale, complex experimental datasets, such as dense biological cell cultures or tissue sections, where the theoretical advantages of global attention over CNNs should become more pronounced. Second, to address the model's sensitivity to propagation distance, future work could explore distance-conditioned network architectures or training regimens that augment the dataset across a continuous range of $z$-distances, potentially culminating in a distance-invariant holographic reconstruction model. Additionally, expanding the framework from 2D planar reconstruction to 3D tomographic imaging remains a significant opportunity, allowing for the volumetric profiling of thick biological samples. Finally, integrating unrolled optimization techniques with the Transformer backbone could further bridge the gap between purely data-driven methods and iterative physics-based solvers.

\section{Conclusion}\label{sec:conclusion}
In this paper, we introduced HoloPASWIN, a physics-aware deep learning framework for single-shot phase retrieval in in-line digital holography. Recognizing the necessity to model global diffraction dependencies, we adapted the Swin Transformer architecture to holographic reconstruction. Coupled with a physics-driven loss function and a residual learning strategy, HoloPASWIN effectively suppresses twin-image artifacts and recovers quantitative phase with high fidelity (average Phase SSIM of 0.986 and Phase PSNR of 46.55 dB on our test set). Through comprehensive ablation and comparative studies, we established the critical roles of complex loss components, physics-consistency constraints, and residual pathways. While simpler geometric datasets may still temporarily favor the inductive biases of classical CNNs, HoloPASWIN provides a fundamentally structured, physics-integrated alternative designed to scale towards the dense entanglements of real-world holographic imaging.

\section*{A1. Derivation of the Reconstructed Field}\label{sec:appendix_derivation}
The twin-image problem arises because the sensor only records intensity. For an inline setup with a uniform reference wave $U_r = 1$, the hologram intensity is:
\begin{equation}
    H(x,y) = |1 + U_o|^2 = 1 + |U_o|^2 + U_o + U_o^*
\end{equation}
Assuming a weakly diffracting object ($|U_o| \ll 1$), we approximate the complex hologram field $\sqrt{H}$ using a first-order Taylor expansion, omitting higher-order terms as their contributions are negligibly small, consistent with the standard weak-object approximation in in-line holography \cite{Huang2024}:
\begin{equation}
    U_{\mathrm{sensor}} \approx \sqrt{1 + U_o + U_o^*} \approx 1 + \frac{1}{2}U_o + \frac{1}{2}U_o^*
\end{equation}
Given that the object wave $U_o$ was forward-propagated by distance $z$ ($\mathcal{P}_z\{O\}$), applying numerical back-propagation by $-z$ yields:
\begin{equation}
    U_{\mathrm{rec}} = \mathcal{P}_{-z} \left\{ U_{\mathrm{sensor}} \right\} \approx \mathcal{P}_{-z}\{1\} + \frac{1}{2}\mathcal{P}_{-z}\{U_o\} + \frac{1}{2}\mathcal{P}_{-z}\{U_o^*\}
\end{equation}
Evaluating each term:
\begin{itemize}
    \item $\mathcal{P}_{-z}\{U_o\} = \mathcal{P}_{-z}\{\mathcal{P}_z\{O\}\} = O$, which is the perfectly focused true object term.
    \item $\mathcal{P}_{-z}\{U_o^*\} = \mathcal{P}_{-z}\{\mathcal{P}_{z}\{O\}^*\} = \mathcal{P}_{-z}\{\mathcal{P}_{-z}\{O^*\}\} = \mathcal{P}_{-2z}\{O^*\}$, which is the unfocused twin-image.
    \item $\mathcal{P}_{-z}\{1\}$ gives the uniform DC background.
\end{itemize}
This demonstrates that standard numerical back-propagation inevitably superimposes the focused object with the out-of-focus twin image.

\section*{Declarations}
\noindent\textbf{Funding}: No external funding was received for this work.

\noindent\textbf{Conflict of Interest}: The authors declare no competing interests.

\noindent\textbf{Data Availability}: The synthetic dataset can be found at Hugging Face \cite{kochmarla2026synthetic_inline_holographical_images_v3}. 

\noindent\textbf{Code Availability}: Source code for this project is hosted at \url{https://github.com/electricalgorithm/holopaswin}.

\bibliography{references}
\end{document}

%% file: architecture.tex
\begin{tikzpicture}
    \tikzset{Box/.pic={\tikzset{/boxblock/.cd,#1}
            \tikzstyle{box}=[every edge/.append style={pic actions, densely dashed, opacity=.7},fill opacity=\opacity, pic actions,fill=\fill]
            
            \pgfmathsetmacro{\y}{\cubey*\scale}
            \pgfmathsetmacro{\z}{\cubez*\scale}
    
            \foreach[count=\i,%
                    evaluate=\i as \xlabel using {array({\boxlabels},\i-1)},%
                    evaluate=\unscaledx as \kx using {\unscaledx*\scale+\prev}, remember=\kx as \prev (initially 0)] 
                    \unscaledx in \cubex
            {
                \pgfmathsetmacro{\x}{\unscaledx*\scale}
                \coordinate (a) at (\kx-\x , \y/2 , \z/2); 
                \coordinate (b) at (\kx-\x ,-\y/2 , \z/2); 
                \coordinate (c) at (\kx    ,-\y/2 , \z/2); 
                \coordinate (d) at (\kx    , \y/2 , \z/2); 
                \coordinate (e) at (\kx    , \y/2 ,-\z/2); 
                \coordinate (f) at (\kx    ,-\y/2 ,-\z/2); 
                \coordinate (g) at (\kx-\x ,-\y/2 ,-\z/2); 
                \coordinate (h) at (\kx-\x , \y/2 ,-\z/2); 
            
                \draw [box] 
                    (d) -- (a) -- (b) -- (c) -- cycle     
                    (d) -- (a) -- (h) -- (e) -- cycle
                    (f) edge (g)
                    (b) edge (g)
                    (h) edge (g)    
                ;
                \path (b) edge ["\xlabel"',midway] (c);
                
                \xdef\LastEastx{\kx} 
            }
            \draw [box] (d) -- (e) -- (f) -- (c) -- cycle; 
            
            \coordinate (a1) at (0 , \y/2 , \z/2);
            \coordinate (b1) at (0 ,-\y/2 , \z/2);
            \tikzstyle{depthlabel}=[pos=0,text width=14*\z,text centered,sloped]       
            \tikzstyle{captionlabel}=[text width=15*\LastEastx/\scale,text centered] 
            
            \path (c) edge ["\small\zlabel"',depthlabel](f); 
            \path (b1) edge ["\ylabel",midway] (a1);  

            \path (\LastEastx/2,-\y/2,+\z/2) + (0,-25pt) coordinate (cap) 
            edge ["\textcolor{black}{ \bf \piccaption}"',captionlabel](cap) ; 
            
            \coordinate (\name-west)   at (0,0,0) ;
            \coordinate (\name-east)   at (\LastEastx, 0,0) ;
            \coordinate (\name-north)  at (\LastEastx/2,\y/2,0);
            \coordinate (\name-south)  at (\LastEastx/2,-\y/2,0);       
            \coordinate (\name-anchor) at (\LastEastx/2, 0,0) ;
            
            \coordinate (\name-near) at (\LastEastx/2,0,\z/2);
            \coordinate (\name-far)  at (\LastEastx/2,0,-\z/2);       
            
            \coordinate (\name-nearwest) at (0,0,\z/2);
            \coordinate (\name-neareast) at (\LastEastx,0,\z/2);
            \coordinate (\name-farwest)  at (0,0,-\z/2);
            \coordinate (\name-fareast)  at (\LastEastx,0,-\z/2);
            
            \coordinate (\name-northeast) at (\name-north-|\name-east);
            \coordinate (\name-northwest) at (\name-north-|\name-west);
            \coordinate (\name-southeast) at (\name-south-|\name-east);
            \coordinate (\name-southwest) at (\name-south-|\name-west);
            
            \coordinate (\name-nearnortheast)  at (\LastEastx, \y/2, \z/2);
            \coordinate (\name-farnortheast)   at (\LastEastx, \y/2,-\z/2);
            \coordinate (\name-nearsoutheast)  at (\LastEastx,-\y/2, \z/2);
            \coordinate (\name-farsoutheast)   at (\LastEastx,-\y/2,-\z/2);
            
            \coordinate (\name-nearnorthwest)  at (0, \y/2, \z/2);
            \coordinate (\name-farnorthwest)   at (0, \y/2,-\z/2);
            \coordinate (\name-nearsouthwest)  at (0,-\y/2, \z/2);
            \coordinate (\name-farsouthwest)   at (0,-\y/2,-\z/2);
            
        },
        /boxblock/.search also={/tikz},
        /boxblock/.cd,
        width/.store        in=\cubex,
        height/.store       in=\cubey,
        depth/.store        in=\cubez,
        scale/.store        in=\scale,
        xlabel/.store       in=\boxlabels,
        ylabel/.store       in=\ylabel,
        zlabel/.store       in=\zlabel,
        caption/.store      in=\piccaption,
        name/.store         in=\name,
        fill/.store         in=\fill,
        opacity/.store      in=\opacity,
        fill={rgb:red,5;green,5;blue,5;white,15},
        opacity=0.4,
        width=2,
        height=13,
        depth=15,
        scale=.2,
        xlabel={{"","","","","","","","","",""}},
        ylabel=,
        zlabel=,
        caption=,
        name=,
    }

    \tikzset{Ball/.pic={\tikzset{/sphere/.cd,#1}	 	

    \pgfmathsetmacro{\r}{\radius*\scale}

    \shade[ball color=\fill,opacity=\opacity] (0,0,0) circle (\r);
    \draw (0,0,0) circle [radius=\r] node[scale=4*\r] {\logo};

    \coordinate (\name-anchor) at ( 0 , 0  , 0) ;
    \coordinate (\name-east)   at ( \r, 0  , 0) ;
    \coordinate (\name-west)   at (-\r, 0  , 0) ;
    \coordinate (\name-north)  at ( 0 , \r , 0) ;
    \coordinate (\name-south)  at ( 0 , -\r, 0) ;

    \path (\name-south) + (0,-20pt) coordinate (caption-node) 
    edge ["\textcolor{black}{\bf \piccaption}"'] (caption-node); 

    },
    /sphere/.search also={/tikz},
    /sphere/.cd,
    radius/.store       in=\radius,
    scale/.store        in=\scale,
    caption/.store      in=\piccaption,
    name/.store         in=\name,
    fill/.store         in=\fill,
    logo/.store         in=\logo,
    opacity/.store      in=\opacity,
    logo=$\Sigma$,
    fill=green,
    opacity=0.10,
    scale=0.2,
    radius=0.5,
    caption=,
    name=,
    }
    \tikzset{RightBandedBox/.pic={\tikzset{/block/.cd,#1}                
            \tikzstyle{box}=[every edge/.append style={pic actions, densely dashed, opacity=.7},fill opacity=\opacity, pic actions,fill=\fill]
            
            \tikzstyle{band}=[every edge/.append style={pic actions, densely dashed, opacity=.7},fill opacity=\bandopacity, pic actions,fill=\bandfill,draw=\bandfill]
            
            \pgfmathsetmacro{\y}{\cubey*\scale}
            \pgfmathsetmacro{\z}{\cubez*\scale}

            \foreach[count=\i,%
                    evaluate=\i as \xlabel using {array({\boxlabels},\i-1)},%
                    evaluate=\unscaledx as \kx using {\unscaledx*\scale+\prev}, remember=\kx as \prev (initially 0)] 
                    \unscaledx in \cubex
            {
                \pgfmathsetmacro{\x}{\unscaledx*\scale}
                \coordinate (a)     at (\kx-\x   , \y/2 , \z/2); 
                \coordinate (art)   at (\kx-\x/3 , \y/2 , \z/2); 
                \coordinate (b)     at (\kx-\x   ,-\y/2 , \z/2); 
                \coordinate (brt)   at (\kx-\x/3 ,-\y/2 , \z/2); 
                \coordinate (c)     at (\kx      ,-\y/2 , \z/2); 
                \coordinate (d)     at (\kx      , \y/2 , \z/2); 
                \coordinate (e)     at (\kx      , \y/2 ,-\z/2); 
                \coordinate (f)     at (\kx      ,-\y/2 ,-\z/2); 
                \coordinate (g)     at (\kx-\x   ,-\y/2 ,-\z/2); 
                \coordinate (h)     at (\kx-\x   , \y/2 ,-\z/2); 
                \coordinate (hrt)   at (\kx-\x/3 , \y/2 ,-\z/2); 
                
                \draw [box] 
                    (d) -- (a) -- (b) -- (c) -- cycle     
                    (d) -- (a) -- (h) -- (e) -- cycle;
                \draw [box]
                    (f) edge (g)
                    (b) edge (g)
                    (h) edge (g);
                \draw [band] 
                    (d) -- (art) -- (brt) -- (c) -- cycle     
                    (d) -- (art) -- (hrt) -- (e) -- cycle;
                \draw [box,fill opacity=0] 
                    (d) -- (a) -- (b) -- (c) -- cycle     
                    (d) -- (a) -- (h) -- (e) -- cycle;            
                    
                \path (b) edge ["\xlabel"',midway] (c);
                
                \xdef\LastEastx{\kx} 
            }
            \draw [box] (d) -- (e) -- (f) -- (c) -- cycle; 
            \draw [band] (d) -- (e) -- (f) -- (c) -- cycle; 
            \draw [pic actions] (d) -- (e) -- (f) -- (c) -- cycle; 
            
            \coordinate (a1) at (0 , \y/2 , \z/2);
            \coordinate (b1) at (0 ,-\y/2 , \z/2);
            \tikzstyle{depthlabel}=[pos=0,text width=14*\z,text centered,sloped]       
            
            \path (c) edge ["\small\zlabels"',depthlabel](f); 
            \path (b1) edge ["\ylabel",midway] (a1);  
            
            \tikzstyle{captionlabel}=[text width=15*\LastEastx/\scale,text centered] 
            \path (\LastEastx/2,-\y/2,+\z/2) + (0,-25pt) coordinate (cap) 
            edge ["\textcolor{black}{ \bf \piccaption}"',captionlabel] (cap); 
            
            \coordinate (\name-west)   at (0,0,0) ;
            \coordinate (\name-east)   at (\LastEastx, 0,0) ;
            \coordinate (\name-north)  at (\LastEastx/2,\y/2,0);
            \coordinate (\name-south)  at (\LastEastx/2,-\y/2,0);       
            \coordinate (\name-anchor) at (\LastEastx/2, 0,0) ;
            
            \coordinate (\name-near) at (\LastEastx/2,0,\z/2);
            \coordinate (\name-far)  at (\LastEastx/2,0,-\z/2);       
            
            \coordinate (\name-nearwest) at (0,0,\z/2);
            \coordinate (\name-neareast) at (\LastEastx,0,\z/2);
            \coordinate (\name-farwest)  at (0,0,-\z/2);
            \coordinate (\name-fareast)  at (\LastEastx,0,-\z/2);
            
            \coordinate (\name-northeast) at (\name-north-|\name-east);
            \coordinate (\name-northwest) at (\name-north-|\name-west);
            \coordinate (\name-southeast) at (\name-south-|\name-east);
            \coordinate (\name-southwest) at (\name-south-|\name-west);
            
            \coordinate (\name-nearnortheast)  at (\LastEastx, \y/2, \z/2);
            \coordinate (\name-farnortheast)   at (\LastEastx, \y/2,-\z/2);
            \coordinate (\name-nearsoutheast)  at (\LastEastx,-\y/2, \z/2);
            \coordinate (\name-farsoutheast)   at (\LastEastx,-\y/2,-\z/2);
            
            \coordinate (\name-nearnorthwest)  at (0, \y/2, \z/2);
            \coordinate (\name-farnorthwest)   at (0, \y/2,-\z/2);
            \coordinate (\name-nearsouthwest)  at (0,-\y/2, \z/2);
            \coordinate (\name-farsouthwest)   at (0,-\y/2,-\z/2);
        },
        /block/.search also={/tikz},
        /block/.cd,
        width/.store        in=\cubex,
        height/.store       in=\cubey,
        depth/.store        in=\cubez,
        scale/.store        in=\scale,
        xlabel/.store       in=\boxlabels,
        ylabel/.store       in=\ylabel,
        zlabel/.store       in=\zlabels,
        caption/.store      in=\piccaption,
        name/.store         in=\name,
        fill/.store         in=\fill,
        bandfill/.store     in=\bandfill,
        opacity/.store      in=\opacity,
        bandopacity/.store  in=\bandopacity,
        fill={rgb:red,5;green,5;blue,5;white,15},
        bandfill={rgb:red,5;green,5;blue,5;white,5},
        opacity=0.4,
        bandopacity=0.6,
        width=2,
        height=13,
        depth=15,
        scale=.2,
        xlabel={{"","","","","","","","","",""}},
        ylabel=,
        zlabel=,
        caption=,
        name=,
    }

    \tikzset{connection/.style={ultra thick,every node/.style={sloped,allow upside down},draw=\edgecolor,opacity=0.7,#1}}
    \tikzset{copyconnection/.style={ultra thick,every node/.style={sloped,allow upside down},draw={rgb:blue,4;red,1;green,1;black,3},opacity=0.7,#1}}

    \node[canvas is zy plane at x=0] (input) at (-3,0,0) {\includegraphics[width=8cm,height=8cm]{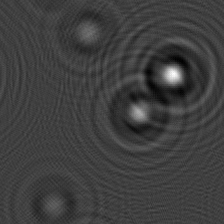}};
    \coordinate (input-east) at (input.east);
    \coordinate (input-west) at (input.west);
    \coordinate (input-north) at (input.north);
    \coordinate (input-south) at (input.south);
    \coordinate (input-anchor) at (input.center);

    \pic[shift={(0,0,0)}] at (0,0,0) {Box={name=asm,caption=ASM,xlabel={{1, }},zlabel=224,fill=\ConvColor,height=40,width=2,depth=40}};
    \draw [connection=] (input-east) -- node {\midarrow} (asm-west);

    \pic[shift={(2,0,0)}] at (asm-east) {Box={name=decomp,caption=Real/Imag Split,xlabel={{2, }},zlabel=224,fill=\ConvColor,height=40,width=4,depth=40}};
    \draw [connection=] (asm-east) -- node {\midarrow} (decomp-west);

    \pic[shift={(2,0,0)}] at (decomp-east) {Box={name=enc1,caption=SWIN Encoder 1,xlabel={{96, }},zlabel=56,fill={rgb:green,0.4;black,0.6},height=20,width=8,depth=20}};
    \draw [connection=] (decomp-east) -- node {\midarrow} (enc1-west);

    \pic[shift={(1.5,0,0)}] at (enc1-east) {Box={name=enc2,caption=SWIN Encoder 2,xlabel={{192, }},zlabel=28,fill={rgb:green,0.4;black,0.6},height=14,width=12,depth=14}};
    \draw [connection=] (enc1-east) -- node {\midarrow} (enc2-west);

    \pic[shift={(1.5,0,0)}] at (enc2-east) {Box={name=enc3,caption=SWIN Encoder 3,xlabel={{384, }},zlabel=14,fill={rgb:green,0.4;black,0.6},height=10,width=18,depth=10}};
    \draw [connection=] (enc2-east) -- node {\midarrow} (enc3-west);

    \pic[shift={(1.5,0,0)}] at (enc3-east) {Box={name=enc4,caption=Bottleneck (SWIN Encoder 4),xlabel={{768, }},zlabel=7,fill={rgb:green,0.2;black,0.8},height=6,width=35,depth=6}};
    \draw [connection=] (enc3-east) -- node {\midarrow} (enc4-west);

    \pic[shift={ (1.5,0,0) }] at (enc4-east) {RightBandedBox={name=dec1,caption=Decoder Upscaler 1/16,xlabel={{ 384, }},zlabel=14,fill={rgb:blue,0.4;black,0.6},bandfill={rgb:white,1;black,2},opacity=0.2,height=10,width=18,depth=10}};
    \draw [connection=] (enc4-east) -- node {\midarrow} (dec1-west);

    \pic[shift={ (1.5,0,0) }] at (dec1-east) {RightBandedBox={name=dec2,caption=Decoder Upscaler 1/8,xlabel={{ 192, }},zlabel=28,fill={rgb:blue,0.6;black,0.4},bandfill={rgb:white,1;black,2},opacity=0.2,height=14,width=12,depth=14}};
    \draw [connection=] (dec1-east) -- node {\midarrow} (dec2-west);

    \pic[shift={ (1.5,0,0) }] at (dec2-east) {RightBandedBox={name=dec3,caption=Decoder Upscaler 1/4,xlabel={{ 96, }},zlabel=56,fill={rgb:blue,0.8;black,0.2},bandfill={rgb:white,1;black,2},opacity=0.2,height=20,width=8,depth=20}};
    \draw [connection=] (dec2-east) -- node {\midarrow} (dec3-west);

    \pic[shift={(2,0,0)}] at (dec3-east) {Box={name=final,caption=Correction Module,xlabel={{2, }},zlabel=224,fill={rgb:blue,0.5;red,0.5;white,0.5},height=40,width=3,depth=40}};
    \draw [connection=] (dec3-east) -- node {\midarrow} (final-west);

    \pic[shift={(2,0,0)}] at (final-east) {Ball={name=sum_res,fill=\SumColor,opacity=0.6,radius=2.0,logo=$+$}};
    \draw [connection=] (final-east) -- node {\midarrow} (sum_res-west);

    \pic[shift={(1.5,0,0)}] at (sum_res-east) {Box={name=out_complex,caption=Clean Complex,xlabel={{2, }},zlabel=224,fill=\ConvColor,height=40,width=3,depth=40}};
    \draw [connection=] (sum_res-east) -- node {\midarrow} (out_complex-west);

    \pic[shift={(1.5,4,0)}] at (out_complex-east) {Box={name=out_amp,caption="",fill={rgb:green,0.8;black,0.2},height=40,width=1,depth=40}};
    \node[canvas is zy plane at x=0] (out_amp_img) at (out_amp-east) [shift={(0.05,0,0)}] {\includegraphics[width=8.0cm,height=8.0cm]{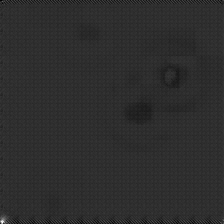}};
    \draw [connection=] (out_complex-east) -- node {\midarrow} (out_amp-west);

    \pic[shift={(1.5,-4,0)}] at (out_complex-east) {Box={name=out_phase,caption=Magnitude and Phase,fill={rgb:blue,0.5;red,0.5;white,0.5},height=40,width=1,depth=40}};
    \node[canvas is zy plane at x=0] (out_phase_img) at (out_phase-east) [shift={(0.05,0,0)}] {\includegraphics[width=8.0cm,height=8.0cm]{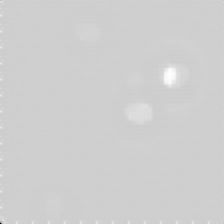}};
    \draw [connection=] (out_complex-east) -- node {\midarrow} (out_phase-west);

    \path (enc3-southeast) -- (enc3-northeast) coordinate[pos=1.25] (enc3-top) ;
    \draw [copyconnection=,dashed] (enc3-northeast) -- node {\copymidarrow}(enc3-top) -- node {\copymidarrow}(dec1-north |- enc3-top) -- node {\copymidarrow} (dec1-north);

    \path (enc2-southeast) -- (enc2-northeast) coordinate[pos=1.25] (enc2-top) ;
    \draw [copyconnection=,dashed] (enc2-northeast) -- node {\copymidarrow}(enc2-top) -- node {\copymidarrow}(dec2-north |- enc2-top) -- node {\copymidarrow} (dec2-north);

    \path (enc1-southeast) -- (enc1-northeast) coordinate[pos=1.25] (enc1-top) ;
    \draw [copyconnection=,dashed] (enc1-northeast) -- node {\copymidarrow}(enc1-top) -- node {\copymidarrow}(dec3-north |- enc1-top) -- node {\copymidarrow} (dec3-north);

    \path (decomp-southeast) -- (decomp-northeast) coordinate[pos=1.25] (decomp-top) ;
    \draw [copyconnection=,] (decomp-northeast) -- node {\copymidarrow}(decomp-top) -- node {\copymidarrow}(sum_res-north |- decomp-top) -- node {\copymidarrow} (sum_res-north);
\end{tikzpicture}

%% file: references.bib
@article{Huang2024,
  title = {Quantitative phase imaging based on holography: trends and new perspectives},
  volume = {13},
  ISSN = {2047-7538},
  url = {http://dx.doi.org/10.1038/s41377-024-01453-x},
  DOI = {10.1038/s41377-024-01453-x},
  number = {1},
  journal = {Light: Science \& Applications},
  publisher = {Springer Science and Business Media LLC},
  author = {Huang,  Zhengzhong and Cao,  Liangcai},
  year = {2024},
  month = jun 
}

@article{park2018,
  title     = "Quantitative phase imaging in biomedicine",
  author    = "Park, Yongkeun and Depeursinge, Christian and Popescu, Gabriel",
  journal   = "Nat. Photonics",
  publisher = "Springer Science and Business Media LLC",
  volume    =  12,
  number    =  10,
  pages     = "578--589",
  month     =  oct,
  year      =  2018,
  language  = "en"
}

@misc{pereira2024review,
      title={A Review of Transformer-Based Models for Computer Vision Tasks: Capturing Global Context and Spatial Relationships}, 
      author={Gracile Astlin Pereira and Muhammad Hussain},
      year={2024},
      eprint={2408.15178},
      archivePrefix={arXiv},
      primaryClass={cs.CV},
      url={https://arxiv.org/abs/2408.15178}, 
}

@article{luo2017receptive,
  title        = "Understanding the effective receptive field in deep
                  convolutional neural networks",
  author       = "Luo, Wenjie and Li, Yujia and Urtasun, Raquel and Zemel,
                  Richard",
  year         =  2017,
  primaryClass = "cs.CV",
  eprint       = "1701.04128"
}

@article{liu2021swin,
  title={Swin transformer: Hierarchical vision transformer using shifted windows},
  author={Liu, Ze and Lin, Yutong and Cao, Yue and Hu, Han and Wei, Yixuan and Zhang, Zheng and Lin, Stephen and Guo, Baining},
  journal={Proceedings of the IEEE/CVF international conference on computer vision},
  pages={10012--10022},
  year={2021}
}

@article{lu2012twin,
  title     = "Twin image elimination from two in-line holograms via phase
               retrieval",
  author    = "Lu Rong, Lu Rong and Feng Pan, Feng Pan and Wen Xiao, Wen Xiao
               and Yan Li, Yan Li and Fanjing Wang, Fanjing Wang",
  journal   = "Chin. Opt. Lett.",
  publisher = "Shanghai Institute of Optics and Fine Mechanics",
  volume    =  10,
  number    =  6,
  pages     = "060902--060904",
  year      =  2012,
  language  = "en"
}

@article{chen2025,
  title     = "Lensless digital holographic reconstruction based on the deep
               unfolding iterative shrinkage thresholding network",
  author    = "Chen, Duofang and Guo, Zijian and Guan, Huidi and Chen, Xueli",
  journal   = "Electronics (Basel)",
  publisher = "MDPI AG",
  volume    =  14,
  number    =  9,
  pages     = "1697",
  month     =  apr,
  year      =  2025,
  copyright = "https://creativecommons.org/licenses/by/4.0/",
  language  = "en"
}

@article{goodman2005introduction,
  author = {Goodman, Joseph W},
  journal = {Introduction to Fourier optics, 3rd ed., by JW Goodman. Englewood, CO: Roberts \& Co. Publishers, 2005},
  title = {Introduction to Fourier optics},
  volume = 1,
  year = 2005
}

@article{latychevskaia2007solution,
  title     = "Solution to the twin image problem in holography",
  author    = "Latychevskaia, Tatiana and Fink, Hans-Werner",
  journal   = "Phys. Rev. Lett.",
  publisher = "American Physical Society (APS)",
  volume    =  98,
  number    =  23,
  pages     = "233901",
  month     =  jun,
  year      =  2007,
  copyright = "http://creativecommons.org/licenses/by/3.0/",
  language  = "en"
}

@misc{kochmarla2026synthetic_inline_holographical_images_v3,
  author       = {Gökhan Koçmarlı},
  title        = {Synthetic Inline Holographical Images v3 (224px Highly Diverse)},
  year         = {2026},
  url          = {https://huggingface.co/datasets/gokhankocmarli/inline-digital-holography-v3},
  note         = {Synthetic dataset for inline holography simulation and reconstruction. Optimized for ViT inputs.}
}

@article{ren2019hrnet,
  title     = "End-to-end deep learning framework for digital holographic
               reconstruction",
  author    = "Ren, Zhenbo and Xu, Zhimin and Lam, Edmund Y",
  journal   = "Adv. Photonics",
  publisher = "SPIE-Intl Soc Optical Eng",
  volume    =  1,
  number    =  01,
  pages     = "1",
  month     =  jan,
  year      =  2019
}

@misc{dosovitskiy2021imageworth16x16words,
      title={An Image is Worth 16x16 Words: Transformers for Image Recognition at Scale}, 
      author={Alexey Dosovitskiy and Lucas Beyer and Alexander Kolesnikov and Dirk Weissenborn and Xiaohua Zhai and Thomas Unterthiner and Mostafa Dehghani and Matthias Minderer and Georg Heigold and Sylvain Gelly and Jakob Uszkoreit and Neil Houlsby},
      year={2021},
      eprint={2010.11929},
      archivePrefix={arXiv},
      primaryClass={cs.CV},
      url={https://arxiv.org/abs/2010.11929}, 
}

@article{rivenson2018phase,
  title={Phase recovery and holographic image reconstruction using deep learning in neural networks},
  author={Rivenson, Yair and Zhang, Yibo and G{\"u}nayd{\i}n, Harun and Teng, Da and Ozcan, Aydogan},
  journal={Light: Science \& Applications},
  volume={7},
  number={2},
  pages={17141--17141},
  year={2018},
  publisher={Nature Publishing Group}
}

@article{wang2020phase,
  title={Phase imaging with an untrained neural network},
  author={Wang, Feng and Bian, Yaoming and Wang, Haowen and Lyu, Meng and Pedrini, Giancarlo and Osten, Wolfgang and Barbastathis, George and Situ, Guohai},
  journal={Light: Science \& Applications},
  volume={9},
  number={1},
  pages={77},
  year={2020},
  publisher={Nature Publishing Group}
}

@ARTICLE{yamaguchi1997,
  title     = "Phase-shifting digital holography",
  author    = "Yamaguchi, I and Zhang, T",
  journal   = "Opt. Lett.",
  publisher = "Optica Publishing Group",
  volume    =  22,
  number    =  16,
  pages     = "1268--1270",
  month     =  aug,
  year      =  1997,
  copyright = "https://doi.org/10.1364/OA\_License\_v1\#VOR",
  language  = "en"
}

@article{liang2021swinir,
  title={SwinIR: Image restoration using swin transformer},
  author={Liang, Jingyun and Cao, Jiezhang and Sun, Guolei and Zhang, Kai and Van Gool, Luc and Timofte, Radu},
  journal={Proceedings of the IEEE/CVF international conference on computer vision},
  pages={1833--1844},
  year={2021}
}

@article{zamir2022restormer,
  title={Restormer: Efficient transformer for high-resolution image restoration},
  author={Zamir, Syed Waqas and Arora, Aditya and Khan, Salman and Hayat, Munawar and Khan, Fahad Shahbaz and Yang, Ming-Hsuan},
  journal={Proceedings of the IEEE/CVF conference on computer vision and pattern recognition},
  pages={5728--5739},
  year={2022}
}
